# Network Coding: Connections Between Information Theory And Estimation Theory

Samah A. M. Ghanem, *Member IEEE*

***Abstract*—In this paper, we prove the existence of fundamental relations between information theory and estimation theory for network-coded flows. When the network is represented by a directed graph $\mathcal{G} = (V, \mathcal{E})$ and under the assumption of uncorrelated noise over information flows between the directed links connecting transmitters, switches (relays), and receivers. We unveil that there yet exist closed-form relations for the gradient of the mutual information with respect to different components of the system matrix M. On the one hand, this result opens a new class of problems casting further insights into effects of the network topology, topological changes when nodes are mobile, and the impact of errors and delays in certain links into the network capacity which can be further studied in scenarios where one source multi-sinks multicasts and multi-source multicast where the invertibility and the rank of matrix M plays a significant role in the decoding process and therefore, on the network capacity. On the other hand, it opens further research questions of finding precoding solutions adapted to the network level.**

## I. INTRODUCTION

When signals are observed in Gaussian noise, there are several intersections between information theory and estimation theory that are related to the measurement system that defines the input-output process. In [1], the authors show that the maximum reliable achievable rate over a wireless link is directly connected to the minimum mean squared error. Later such relations were derived for linear vector Gaussian channels in [2], for multiple access channels (MAC) in [3], and [4], for signals with general distributions. Recently, in [5], and [6] the author unveiled a generalized fundamental relation between the mutual information and the minimum mean squared error which applies to multiuser Gaussian channels or - on the network terminology - to *network cuts*. This has motivated the investigation of the interplay between information theory and estimation theory on a network level. In particular, when network coding is considered and information flows are decoded and recoded over the transmission chain, it is of particular importance to address such connections. The problem can be tackled from a rate distortion perspective when channels are noiseless [7]. However, to have an understanding of the propagation of errors over the wireless network and the effects of the topology, it is of particular relevance to revisit the problem on the network level. In particular, we need to investigate the network capacity with respect to the minimum mean squared error at the receiver side under the assumption of uncorrelated noisy channels. To the best of the author's knowledge, this work is the first in the literature to address such problem.

The main contribution of this paper is that we proved there yet exist connections between information theory and estimation theory in noisy coded networks. Such connections provides an abstraction of the wireless channel, where optimal designs of communication systems and optimal methods can be corroborated. Particularly, resource allocation, scheduling, precoding, and decoding, etc. can be addressed from a network level perspective taking into account the possible dimensions that can be exploited by having designs adapted to the awareness of the physical layer and the network topology.

The remainder of the paper is organized as follows, section II introduces the system model. Section III introduces fundamental connections between information theory and estimation theory in a network level. Section IV introduces some analysis. Section V concludes the paper.

The following notation is employed, upper case letters denote matrices, lower case denotes vectors, the superscript $(.)^{\mathrm{T}}$, and $(.)^{\dagger}$ denote transpose and conjugate transpose operations. $(.)^{*}$ denotes optimum, $\mathrm{Tr}\{.\}$ denotes the trace of a matrix, $\mathbb{E}(.)$ denotes the expectation. $\|\mathbf{X}\| = \sqrt{(\mathrm{Tr}\{\mathbf{X}\mathbf{X}^{\dagger}\})}$ which reduces to the $\mathrm{L}^2$-norm $\|\mathbf{x}\|$ in the special case of a vector. And $\nabla_{\boldsymbol{X}}$ denotes the gradient with respect to $\mathbf{X}$.

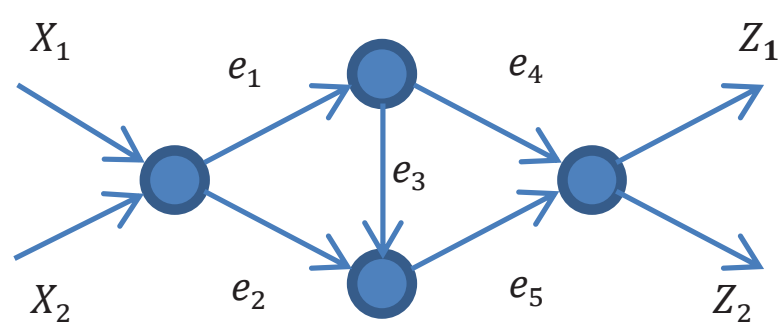

Figure 1. A network topology with once source one sink multicast is given by $\mathcal{N} = (v_1, v_4, X(v_1, 1), X(v_1, 2), X(v_1, 3))$. $\{X_1, X_2\}$ are the input vectors of multicast packets. $\{e_1, e_2, e_3, e_4, e_5\}$ are the set of edges (links). $\{Z_1, Z_2\}$ are the output vectors.

## II. SYSTEM MODEL

A communication network is a collection of directed links connecting transmitters, switches (relays), and receivers [8]. It may be represented by a directed graph $\boldsymbol{\mathcal{G}} = (\boldsymbol{V}, \boldsymbol{\mathcal{E}})$ with a vertex set $\boldsymbol{V}$ and an edge (link) set $\boldsymbol{\mathcal{E}}$. The source nodes are represented by $\{\boldsymbol{v_1}, \boldsymbol{v_2}, \dots, \boldsymbol{v_N}\} \subseteq \boldsymbol{V}$, and the sink nodes are $\{\boldsymbol{u_1}, \boldsymbol{u_2}, \dots, \boldsymbol{u_N}\} \subseteq \boldsymbol{V}$. Therefore, we can formulate the linear network matrix representation as follows,

$$\boldsymbol{z} = \mathbf{M}\boldsymbol{x} + \boldsymbol{n} \tag{1}$$

With input vector of discrete random processes observable at source node $\boldsymbol{v}$ is $\boldsymbol{x}^T = \{X(\boldsymbol{v},\mathbf{1}), X(\boldsymbol{v},\mathbf{2}), \dots, X(\boldsymbol{v},\boldsymbol{\mu}(\boldsymbol{v}))\}$, output vector $\boldsymbol{z}^T = \{Z(\boldsymbol{u},\mathbf{1}), Z(\boldsymbol{u},\mathbf{2}), \dots, Z(\boldsymbol{u},\boldsymbol{v}(\boldsymbol{u}))\}$, with $\mathbf{M} = \mathbf{AGB}$ is the system matrix, $\boldsymbol{n}$ is the vector of uncorrelated random Gaussian noise. Therefore, for the special network of Figure 1, we can rewrite (1) as follows,

$$\begin{bmatrix} Z_1 \\ Z_2 \end{bmatrix} = \underbrace{\begin{bmatrix} \gamma_{e4,1} & \gamma_{e5,1} \\ \gamma_{e4,2} & \gamma_{e5,2} \end{bmatrix}}_{\mathbf{A}} \underbrace{\begin{bmatrix} \beta_{e1,e4} & \mathbf{0} \\ \beta_{e1,e3}\beta_{e3,e5} & \beta_{e2,e5} \end{bmatrix}}_{\mathbf{G}} \underbrace{\begin{bmatrix} \alpha_{1,e1} & \alpha_{1,e2} \\ \alpha_{2,e1} & \alpha_{2,e2} \end{bmatrix}}_{\mathbf{B}} \begin{bmatrix} X_1 \\ X_2 \end{bmatrix} + \begin{bmatrix} N_1 \\ N_2 \end{bmatrix} \quad (2)$$

Notice that the matrices **B**, **G**, and **A** are derived from the data flow across the network topology. For instance, the non-existence of a two-way link between the intermediate nodes in Figure 1 appears as a zero element in the matrix **G** which is also called the topology matrix.

## III. Connections between Inf. Theory and Est. Theory in Noisy Coded Networks

The network capacity $\boldsymbol{\mathcal{C}}$ is defined by the max-flow min cut theorem which is known to be achieved by coding over network flows [9]. However, the information theoretic definition of capacity is the maximum mutual information between the input-output processes on a network level, which can be written as follows,

$$\boldsymbol{\mathcal{C}} = max\, \boldsymbol{I}(\boldsymbol{x};\boldsymbol{z}) \quad (3)$$

We will not define any constraints here, however, our aim is to derive connections between information measures and estimation measures for uncorrelated noisy channels on a network level[1]. Therefore, our aim is to find a closed form relations between the network capacity and the minimum mean squared error at the sink side. The following theorem provides a proof of existence of connections between information theory and estimation theory in noisy coded networks. The implication of such result is multifold. In particular, one of the fundamental outcomes is that we can build up a backward traceability to where the error occurs on the chain of transmission. Therefore, we allow reliable recovery, better resource allocation and resource planning.

***Theorem 1*****:** The relation between the gradient of the mutual information and the non-linear MMSE with respect to the decoding matrix **A**, topology matrix **G** , and precoding matrix **B**, for the linear network in (1) satisfies,

$$\boldsymbol{\nabla_A I}(\boldsymbol{x};\boldsymbol{z}) = \mathbf{MEB}^\dagger\mathbf{G}^\dagger = \mathbf{AGBEB}^\dagger\mathbf{G}^\dagger \quad (4)$$
$$\boldsymbol{\nabla_G I}(\boldsymbol{x};\boldsymbol{z}) = \mathbf{A}^\dagger\,\mathbf{MEB}^\dagger = \mathbf{A}^\dagger\mathbf{AGBEB}^\dagger \quad (5)$$
$$\boldsymbol{\nabla_B I}(\boldsymbol{x};\boldsymbol{z}) = \mathbf{G}^\dagger\mathbf{A}^\dagger\mathbf{ME} = \mathbf{G}^\dagger\mathbf{A}^\dagger\mathbf{AGBE} \quad (6)$$

With the network error matrix is given by,
$\mathbf{E} = \mathbb{E}((\boldsymbol{x}-\widehat{\boldsymbol{x}})(\boldsymbol{x}-\widehat{\boldsymbol{x}})^\dagger)$

and the conditional mean estimator of the input flow vector at the source given the output flow vector at the sink is given by,
$\widehat{\boldsymbol{x}} = \mathbb{E}(\boldsymbol{x}|\boldsymbol{z}) = \sum_x \frac{x\, p(\boldsymbol{Z}|\boldsymbol{X})p(x)}{p(z)}$

*Proof: See Appendices A, B, and C*

[1]Note that we assume that the noise is uncorrelated along the information flow. However, yet the network coding mechanism enforces correlation within the information flow encoding/re-encoding process at each node, not applied to the noise.

Notice that the derived relation between the gradient of the mutual information and the MMSE in Theorem 1 will lead to a new formulation of $\mathbb{E}(\boldsymbol{x}|\boldsymbol{z})$, with respect to the precoding, topology, and decoding matrices, as provided in the following Theorem.

***Theorem 2*****:** The estimates of the input vector $\boldsymbol{x}$ of the linear network with one source one sink multicast model in (1) given the output $\boldsymbol{z}$ can be expressed as,

$$\mathbb{E}(\boldsymbol{x}|\boldsymbol{z}) = [\boldsymbol{x} + \boldsymbol{\nabla_z} \log p_z(\boldsymbol{z})]\mathbf{B}^{-1}(\mathbf{I}-\mathbf{F})\mathbf{A}^{-1} \quad (7)$$

*Proof: See Appendix D*

Theorem 2 is of particular importance because it shows directly that the decoding process doesn't require that the matrix **G** to be invertible. In fact, we need the system matrix **M** to be invertible. Therefore, a projection of the matrix $\mathbf{G} = (\mathbf{I}-\mathbf{F})^{-1}$ suffice to restore the information flow where the projection is $\mathbf{G}^{-1} = \mathbf{I} - \mathbf{F}$. Or in other words, the optimal topology is the one that will let us have **M** to be invertible. Worth to notice that the probability of the output is bounded as $0 \le p_z(\boldsymbol{z}) \le 1$, in turn, the information flow estimate is bounded by,

$$[\boldsymbol{x}+\mathbf{1}]\mathbf{B}^{-1}(\mathbf{I}-\mathbf{F})\mathbf{A}^{-1} \ge \mathbb{E}(\boldsymbol{x}|\boldsymbol{z}) \ge \boldsymbol{x}\mathbf{B}^{-1}(\mathbf{I}-\mathbf{F})\mathbf{A}^{-1} \quad (8)$$

We can further simplify (8) to obtain, $\boldsymbol{x} \le \mathbb{E}(\boldsymbol{x}|\boldsymbol{z})\,\mathbf{AGB} \le [\boldsymbol{x}+\mathbf{1}]$, where $\mathbf{B}^{-1}\mathbf{G}^{-1}\mathbf{A}^{-1}$ plays the role of an inverse filter in the estimation process at the sink which minimizes the distance between the information flow and their nonlinear estimates, i.e., minimizes the network MMSE.

## IV. Analysis

In this section, we analyze the setup from an error propagation perspective and also from the perspective of the network capacity changes with respect to the arbitrary parameters of the network. In particular, if some errors occur at certain bits of the transmitted packet, while the intermediate nodes need to code across the packets, the bits which have already flipped can be decoded correctly at the sink, if the nodes utilize decode/recode and forward, preserving the flow over the paths. Therefore, it's very important to notice that Theorem 1 classifies the change in the network capacity in the direction of different parameters of the network, precoding, topology, or decoding.

***Example 1***

Let us do analysis for the most interesting part, which is the topology, and test some changes in the topology adding or removing a link. The gradient of the mutual information with respect to the topology matrix **G** for the network[2] in Figure 1 and based on Theorem 1 (6) can be written as,

$$\boldsymbol{\nabla_G I}(\boldsymbol{x};\boldsymbol{z}) = \mathbf{A}^\dagger\,\mathbf{MEB}^\dagger = \mathbf{A}^\dagger\mathbf{AGBEB}^\dagger = \begin{bmatrix} \boldsymbol{\psi}_{11} & \boldsymbol{\psi}_{12} \\ \boldsymbol{\psi}_{21} & \boldsymbol{\psi}_{22} \end{bmatrix} \quad (9)$$

For the network in Figure 1, due to the lack of space, we take the first element in the matrix (9) to analyze,

[2]Note that the problem will be much more complicated if multiple flows are considered to be arriving from multiple sources to multiple sinks. This setup may evolve a sparse structure of the network model which is out of the scope of this work.

$$
\begin{aligned}
\psi_{11} = \mathbf{E}_{11}\big(&\gamma_{e4,1}{}^2\beta_{e1,e4}\alpha_{1,e1}{}^2 + \gamma_{e4,1}\gamma_{e5,1}\beta_{e1,e3}\beta_{e3,e5}\alpha_{1,e1}{}^2 \\
&+ \gamma_{e4,1}\gamma_{e5,1}\beta_{e2,e5}\alpha_{2,e1}\alpha_{1,e1} + \gamma_{e4,2}{}^2\beta_{e1,e4}\alpha_{1,e1}{}^2 \\
&+ \gamma_{e4,2}\gamma_{e5,2}\beta_{e1,e3}\beta_{e3,e5}\alpha_{1,e1}{}^2 \\
&+ \gamma_{e4,1}\gamma_{e5,2}\beta_{e2,e5}\alpha_{2,e1}\alpha_{1,e1}\big) \\
+\mathbf{E}_{12}(&\gamma_{e4,1}{}^2\beta_{e1,e4}\alpha_{1,e1}\alpha_{1,e2} + \gamma_{e4,1}\gamma_{e5,1}\beta_{e1,e3}\beta_{e3,e5}\alpha_{1,e1}\alpha_{1,e2} \\
&+ \gamma_{e4,1}\gamma_{e5,1}\beta_{e2,e5}\alpha_{2,e1}\alpha_{1,e2} + \gamma_{e4,2}{}^2\beta_{e1,e4}\alpha_{1,e1}\alpha_{1,e2} \\
&+ \gamma_{e4,2}\gamma_{e5,2}\beta_{e1,e3}\beta_{e3,e5}\alpha_{1,e1}\alpha_{1,e2} \\
&+ \gamma_{e4,2}\gamma_{e5,2}\beta_{e2,e5}\alpha_{2,e1}\alpha_{1,e2}) \\
+\mathbf{E}_{21}(&\gamma_{e4,1}{}^2\beta_{e1,e4}\alpha_{1,e1}\alpha_{1,e2} + \gamma_{e4,1}\gamma_{e5,1}\beta_{e1,e3}\beta_{e3,e5}\alpha_{1,e2}\alpha_{1,e1} \\
&+ \gamma_{e4,1}\gamma_{e5,1}\beta_{e2,e5}\alpha_{2,e2}\alpha_{1,e1} + \gamma_{e4,2}{}^2\beta_{e1,e4}\alpha_{1,e1}\alpha_{1,e2} \\
&+ \gamma_{e4,2}\gamma_{e5,2}\beta_{e1,e3}\beta_{e3,e5}\alpha_{1,e1}\alpha_{1,e2} \\
&+ \gamma_{e4,2}\gamma_{e5,2}\beta_{e2,e5}\alpha_{2,e2}\alpha_{1,e1}) \\
+E_{22}(&\gamma_{e4,1}{}^2\beta_{e1,e4}\alpha_{1,e2}{}^2 + \gamma_{e4,1}\gamma_{e5,1}\beta_{e1,e3}\beta_{e3,e5}\alpha_{1,e2}{}^2 \\
&+ \gamma_{e4,1}\gamma_{e5,1}\beta_{e2,e5}\alpha_{2,e2}\alpha_{1,e2} + \gamma_{e4,2}{}^2\beta_{e1,e4}\alpha_{1,e2}{}^2 \\
&+ \gamma_{e4,2}\gamma_{e5,2}\beta_{e1,e3}\beta_{e3,e5}\alpha_{1,e2}{}^2 \\
&+ \gamma_{e4,2}\gamma_{e5,2}\beta_{e2,e5}\alpha_{2,e2}\alpha_{1,e2})
\end{aligned}
$$

Assume that the topology changes such that the edge *e3* is disconnected, therefore the topology matrix is diagonal, and we will directly see that the error will not propagate through this link so that the change in the capacity will be in the direction of the existing new topology. Therefore, we can write the first element of the gradient in (9) as follows,

$$
\begin{aligned}
\psi_{11} = \mathbf{E}_{11}\big(&\gamma_{e4,1}{}^2\beta_{e1,e4}\alpha_{1,e1}{}^2 + \gamma_{e4,1}\gamma_{e5,1}\beta_{e2,e5}\alpha_{2,e1}\alpha_{1,e1} \\
&+ \gamma_{e4,2}{}^2\beta_{e1,e4}\alpha_{1,e1}{}^2 + \gamma_{e4,1}\gamma_{e5,2}\beta_{e2,e5}\alpha_{2,e1}\alpha_{1,e1}\big) \\
+\mathbf{E}_{12}(&\gamma_{e4,1}{}^2\beta_{e1,e4}\alpha_{1,e1}\alpha_{1,e2} + \gamma_{e4,1}\gamma_{e5,1}\beta_{e2,e5}\alpha_{2,e1}\alpha_{1,e2} \\
&+ \gamma_{e4,2}{}^2\beta_{e1,e4}\alpha_{1,e1}\alpha_{1,e2} + \gamma_{e4,2}\gamma_{e5,2}\beta_{e2,e5}\alpha_{2,e1}\alpha_{1,e2}) \\
+\mathbf{E}_{21}(&\gamma_{e4,1}{}^2\beta_{e1,e4}\alpha_{1,e1}\alpha_{1,e2} + \gamma_{e4,1}\gamma_{e5,1}\beta_{e2,e5}\alpha_{2,e2}\alpha_{1,e1} \\
&+ \gamma_{e4,2}{}^2\beta_{e1,e4}\alpha_{1,e1}\alpha_{1,e2} + \gamma_{e4,2}\gamma_{e5,2}\beta_{e2,e5}\alpha_{2,e2}\alpha_{1,e1}) \\
+E_{22}(&\gamma_{e4,1}{}^2\beta_{e1,e4}\alpha_{1,e2}{}^2 + \gamma_{e4,1}\gamma_{e5,1}\beta_{e2,e5}\alpha_{2,e2}\alpha_{1,e2} + \gamma_{e4,2}{}^2\beta_{e1,e4}\alpha_{1,e2}{}^2 \\
&+ \gamma_{e4,2}\gamma_{e5,2}\beta_{e2,e5}\alpha_{2,e2}\alpha_{1,e2})
\end{aligned}
$$

If we lose the edges $\{e2, e5\}$, the first element of the gradient in (9) will be as follows,

$$
\begin{aligned}
\psi_{11} = \mathbf{E}_{11}\big(&\gamma_{e4,1}{}^2\beta_{e1,e4}\alpha_{1,e1}{}^2 + \gamma_{e4,2}{}^2\beta_{e1,e4}\alpha_{1,e1}{}^2\big) \\
&+ \mathbf{E}_{12}(\gamma_{e4,1}{}^2\beta_{e1,e4}\alpha_{1,e1}\alpha_{1,e2} + \gamma_{e4,2}{}^2\beta_{e1,e4}\alpha_{1,e1}\alpha_{1,e2}) \\
&+ \mathbf{E}_{21}(\gamma_{e4,1}{}^2\beta_{e1,e4}\alpha_{1,e1}\alpha_{1,e2} + \gamma_{e4,2}{}^2\beta_{e1,e4}\alpha_{1,e1}\alpha_{1,e2}) \\
&+ E_{22}(\gamma_{e4,1}{}^2\beta_{e1,e4}\alpha_{1,e2}{}^2 + \gamma_{e4,2}{}^2\beta_{e1,e4}\alpha_{1,e2}{}^2)
\end{aligned}
$$

To finalize the discussion about the change of the capacity with respect to topological changes, we can observe that the elements of the gradient of the mutual information produce four different equations with respect to the error and other network parameters. Therefore, we can optimize one or more parameters of the system given a certain design criterion we aim to.

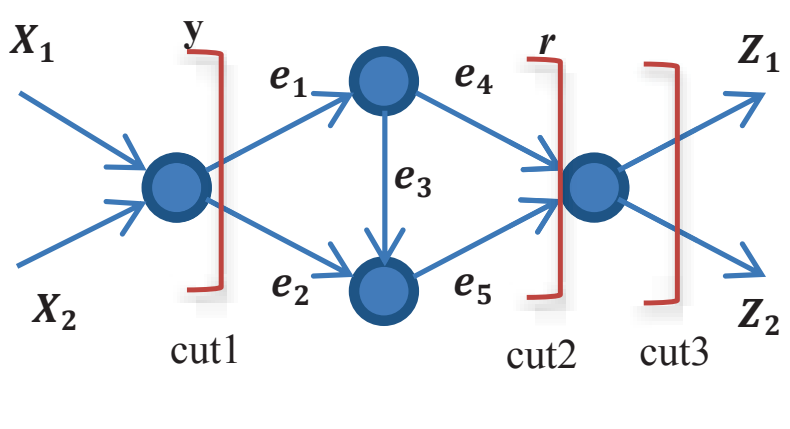

Figure 2. A network topology with once source one sink multicast with cuts at vertex $v_1$ and vertex $v_2$. Network cut1 corresponds to $\boldsymbol{y} = \mathbf{B}\boldsymbol{x} + \tilde{\boldsymbol{n}}$, cut2 corresponds to $\boldsymbol{r} = \mathbf{GB}\boldsymbol{x} + \tilde{\boldsymbol{n}}$, and cut3 corresponds to $\boldsymbol{z} = \mathbf{M}\boldsymbol{x} + \boldsymbol{n}$.

### *Example 2*

Suppose that we need to look at the network in an approach similar to the max-flow min-cut theorem, where we will cut the network into sections and see how the network capacity evolves with respect to each addition, we have a cut at first node or vertex $v_1$, then the network model will be represented as,

$$\boldsymbol{y} = \mathbf{B}\boldsymbol{x} + \tilde{\boldsymbol{n}} \tag{10}$$

However, if we cut the network before the sink at vertex $v_4$, then the network model will be represented as,

$$\boldsymbol{r} = \mathbf{GB}\boldsymbol{x} + \tilde{\boldsymbol{n}} \tag{11}$$

If we cut at the sink side, the network model will be complete as in (1). $\tilde{\boldsymbol{n}}$ and $\tilde{\boldsymbol{n}}$ are vectors of uncorrelated random Gaussian noise. Therefore, if the decode/recode and forward nodes restore exactly the same flow, along a topology of symmetric links that could preserve the trace or the eigenvalues of the network representation, then the error will be the same[3]. In particular, if the gradient of the mutual information -defined over the input-output of a network- with respect to some arbitrary parameters is a function of $\boldsymbol{E}$, the error matrix will be the same wherever the cut happened in the network. The following theorem specializes the result of Theorem 1 into special cuts in the network. Figure 2 illustrates the concept we are trying to prove.

***Theorem 3*****:** The relation between the gradient of the mutual information and the non-linear MMSE with respect to the precoding matrix $\mathbf{B}$, and topology matrix $\mathbf{G}$ for the linear network in (10) and (11) satisfies,

$$\nabla_{B} \boldsymbol{I}(\boldsymbol{x};\boldsymbol{y}) = \mathbf{B}\tilde{\mathbf{E}} \tag{12}$$

$$\nabla_{B} \boldsymbol{I}(\boldsymbol{x};\boldsymbol{r}) = \mathbf{G}^{\dagger}\mathbf{GB}\breve{\mathbf{E}} \tag{13}$$

$$\nabla_{G} \boldsymbol{I}(\boldsymbol{x};\boldsymbol{r}) = \mathbf{GB}\breve{\mathbf{E}}\mathbf{B}^{\dagger} \tag{14}$$

With the error matrices are defined as,

$\tilde{\mathbf{E}} = \mathbb{E}\big((\boldsymbol{x} - \hat{\ddot{\boldsymbol{x}}})(\boldsymbol{x} - \hat{\ddot{\boldsymbol{x}}})^{\dagger}\big)$ and, $\breve{\mathbf{E}} = \mathbb{E}\big((\boldsymbol{x} - \breve{\ddot{\boldsymbol{x}}})(\boldsymbol{x} - \breve{\ddot{\boldsymbol{x}}})^{\dagger}\big)$ respectively, and their corresponding conditional mean estimator of the input flow vector at the source given the output flow vector at the sink is given respectively as,

$\hat{\ddot{\boldsymbol{x}}} = \mathbb{E}(\boldsymbol{x}|\boldsymbol{y}) = \sum_x \frac{x\, p(\boldsymbol{y}|\boldsymbol{x}) p(x)}{p(\boldsymbol{y})}$ and, $\breve{\ddot{\boldsymbol{x}}} = \mathbb{E}(\boldsymbol{x}|\boldsymbol{r}) = \sum_x \frac{x\, p(\boldsymbol{r}|\boldsymbol{x}) p(x)}{p(\boldsymbol{r})}$

*Proof: following the same steps of the proof of Theorem 1.*

The implication of Theorem 3 is of great importance to see that the abstraction of a network includes the physical layer effects. Suppose that the precoding matrix $\mathbf{B}$ corresponds to a channel matrix, we can rewrite the model in (10) as follows,

$$\begin{bmatrix} y1 \\ y2 \end{bmatrix} = \underbrace{\begin{bmatrix} \alpha_{1,e1} & \alpha_{1,e2} \\ \alpha_{2,e1} & \alpha_{2,e2} \end{bmatrix}}_{B} \begin{bmatrix} X1 \\ X2 \end{bmatrix} + \begin{bmatrix} \tilde{n}1 \\ \tilde{n}2 \end{bmatrix} \tag{15}$$

Where the system corresponds to a MIMO channel model with a channel that aligns transmit directions given a diagonal unit norm normalized transmitted power, thus, we can rewrite (15) as follows,

$$\begin{bmatrix} y1 \\ y2 \end{bmatrix} = \underbrace{\begin{bmatrix} h_{1,1} & h_{1,2} \\ h_{2,1} & h_{2,2} \end{bmatrix}}_{\mathbf{H}} \begin{bmatrix} X1 \\ X2 \end{bmatrix} + \begin{bmatrix} \tilde{n}1 \\ \tilde{n}2 \end{bmatrix} \tag{16}$$

[3]Analysis of error propagation and backward traceability of the error along the network topology is a topic of interest for future research.

## V. CONCLUSIONS

In this paper, we prove the existence of intersections between information theory and estimation theory on a network level. The gradient of the mutual information which corresponds to the change of the network capacity in the direction of precoding, decoding, and topological matrix is a function with respect to those arbitrary parameters and the error that takes place in different links in the network. This contribution is of fundamental importance to future designs of wireless networks with topology awareness. In addition it is of particular relevance to address problems like adaptive resource allocation and flow association based on maximum flows under wireless channel changes, i.e., with physical layer awareness. As we have highlighted in the analysis, the most abstraction of the capacity is a network capacity, and the least granularity is the capacity of channel with multiple input multiple output (MIMO) model. Future research will consider precoding on a network level and the problem of multi-source multi-sink networks.

## APPENDIX A

### Proof of Theorem 1-Part I

The conditional probability density for the one source-one sink multicast network can be written as follows,

$$p_{z|x}(\boldsymbol{z}|\boldsymbol{x}) = \frac{1}{\pi^{n_r}} e^{-\|\boldsymbol{z}-\mathbf{M}\boldsymbol{x}\|^2}$$

Thus, the corresponding mutual information is,

$$I(\boldsymbol{x};\boldsymbol{z}) = \mathbb{E}\left(log \frac{p(\boldsymbol{Z}|\boldsymbol{X})}{p(\boldsymbol{z})}\right)$$
$$= -n_r\, log(\pi e) - \mathbb{E}(log\, p(\boldsymbol{z}))$$
$$= -n_r\, log(\pi e) - \int p(\boldsymbol{z})\, log\, p(\boldsymbol{z}) d\boldsymbol{z}$$

Then,

$$\frac{\partial I(\boldsymbol{x};\boldsymbol{z})}{\partial \mathbf{A}^\dagger} = -\frac{\partial}{\partial \mathbf{A}^\dagger}\int p(\boldsymbol{z})\, log\, p(\boldsymbol{z}) d\boldsymbol{z}$$
$$= -\int \left(p(\boldsymbol{z})\frac{1}{p(\boldsymbol{z})} + log\, p(\boldsymbol{z})\right)\frac{\partial p(\boldsymbol{z})}{\partial \mathbf{A}^\dagger} d\boldsymbol{z}$$
$$= -\int (1 + log\, p(\boldsymbol{z}))\frac{\partial p(\boldsymbol{z})}{\partial \mathbf{A}^\dagger} d\boldsymbol{z}$$

Where,

$$p(\boldsymbol{z}) = \sum_x p(\boldsymbol{x}) p_{z|x}(\boldsymbol{z}|\boldsymbol{x}) = \mathbb{E}_x\left(p_{z|x}(\boldsymbol{z}|\boldsymbol{x})\right)$$

The derivative of the conditional output can be written as,

$$\frac{\partial p_{z|x}(\boldsymbol{z}|\boldsymbol{x})}{\partial \mathbf{A}^\dagger}$$
$$= -p_{z|x}(\boldsymbol{z}|\boldsymbol{x})\frac{\partial}{\partial \mathbf{A}^\dagger}(\boldsymbol{z}-\mathbf{M}\boldsymbol{x})^\dagger(\boldsymbol{z}-\mathbf{M}\boldsymbol{x})$$
$$= -p_{z|x}(\boldsymbol{z}|\boldsymbol{x})\frac{\partial}{\partial \mathbf{A}^\dagger}(\boldsymbol{z}-\mathbf{AGB}\boldsymbol{x})^\dagger(\boldsymbol{z}-\mathbf{AGB}\boldsymbol{x})$$
$$= p_{z|x}(\boldsymbol{z}|\boldsymbol{x})(\boldsymbol{z}-\mathbf{M}\boldsymbol{x})\boldsymbol{x}^\dagger\mathbf{B}^\dagger\mathbf{G}^\dagger$$
$$= \nabla_z p_{z|x}(\boldsymbol{z}|\boldsymbol{x})\boldsymbol{x}^\dagger\mathbf{B}^\dagger\mathbf{G}^\dagger$$

Therefore we have,

$$\mathbb{E}_x\left(\nabla_{A^\dagger} p_{z|x}(\boldsymbol{z}|\boldsymbol{x})\right) = \mathbb{E}_x\left(\nabla_z p_{z|x}(\boldsymbol{z}|\boldsymbol{x})\boldsymbol{x}^\dagger\mathbf{B}^\dagger\mathbf{G}^\dagger\right)$$
$$\frac{\partial I(\boldsymbol{x};\boldsymbol{z})}{\partial \mathbf{A}^\dagger} = \int (1 + log\, p(\boldsymbol{z}))\,\mathbb{E}_x\left(\nabla_z p_{z|x}(\boldsymbol{z}|\boldsymbol{x})\right)\boldsymbol{x}^\dagger\mathbf{B}^\dagger\mathbf{G}^\dagger d\boldsymbol{z}$$
$$= \mathbb{E}\left(\left(\int (1 + log\, p(\boldsymbol{z}))\,\nabla_z p_{z|x}(\boldsymbol{z}|\boldsymbol{x}) d\boldsymbol{z}\right)\boldsymbol{x}^\dagger\mathbf{B}^\dagger\mathbf{G}^\dagger\right)$$

Using integration by parts applied to the real and imaginary parts of $\boldsymbol{z}$,

$$\int (1 + log\,_e p(\boldsymbol{z}))\frac{\partial p_{z|x}(\boldsymbol{z}|\boldsymbol{x})}{\partial \boldsymbol{t}} dt$$
$$= \int (1 + log\,_e p(\boldsymbol{z}))\, p_{z|x}(\boldsymbol{z}|\boldsymbol{x})|_{-\infty}^{\infty}$$
$$- \int_{-\infty}^{\infty}\frac{1}{p(\boldsymbol{z})}\frac{\partial p(\boldsymbol{z})}{\partial \boldsymbol{t}}\, p_{z|x}(\boldsymbol{z}|\boldsymbol{x}) dt$$

The first term goes to zero as $\|\boldsymbol{z}\| \xrightarrow{yields} \infty$, Then,

$$\frac{\partial I(\boldsymbol{x};\boldsymbol{z})}{\partial \mathbf{A}^\dagger} = \mathbb{E}\left(\left(-\int \frac{p_{z|x}(\boldsymbol{z}|\boldsymbol{x})}{p(\boldsymbol{z})}\nabla_z p(\boldsymbol{z})\, d\boldsymbol{z}\right)\boldsymbol{x}^\dagger\mathbf{B}^\dagger\mathbf{G}^\dagger\right)$$
$$= -\int \nabla_z p(\boldsymbol{z})\mathbb{E}_x\left(\frac{p_{z|x}(\boldsymbol{z}|\boldsymbol{x})}{p(\boldsymbol{z})}\boldsymbol{x}^\dagger\mathbf{B}^\dagger\mathbf{G}^\dagger\right) d\boldsymbol{z}$$
$$= -\int \nabla_z p(\boldsymbol{z})\,\mathbb{E}(\boldsymbol{x}|\boldsymbol{z})^\dagger\,\mathbf{B}^\dagger\mathbf{G}^\dagger d\boldsymbol{z}$$

However,

$$\nabla_z p(\boldsymbol{z}) = \nabla_z \mathbb{E}_x\left(p_{z|x}(\boldsymbol{z}|\boldsymbol{x})\right) = \mathbb{E}_x\left(\nabla_z p_{z|x}(\boldsymbol{z}|\boldsymbol{x})\right)$$
$$= -\mathbb{E}_x\left(p_{z|x}(\boldsymbol{z}|\boldsymbol{x})(\boldsymbol{z}-\mathbf{M}\boldsymbol{x})\right)$$
$$= -\mathbb{E}_x(p_z(\boldsymbol{z})(\boldsymbol{z}-\mathbf{M}\boldsymbol{x})|\boldsymbol{z})$$
$$= -p_z(\boldsymbol{z})\left(\boldsymbol{z}-\mathbf{M}\mathbb{E}(\boldsymbol{x}|\boldsymbol{z})\right)$$

Thus,

$$\frac{\partial I(\boldsymbol{x};\boldsymbol{z})}{\partial \mathbf{A}^\dagger} = \int p_z(\boldsymbol{z})\left(\boldsymbol{z}-\mathbf{M}\mathbb{E}(\boldsymbol{x}|\boldsymbol{z})\right)\mathbb{E}(\boldsymbol{x}|\boldsymbol{z})^\dagger\,\mathbf{B}^\dagger\mathbf{G}^\dagger d\boldsymbol{z}$$
$$= \mathbb{E}(\boldsymbol{z}\boldsymbol{x}^\dagger)\mathbf{B}^\dagger\mathbf{G}^\dagger - \mathbb{E}(\mathbf{M}\mathbb{E}(\boldsymbol{x}|\boldsymbol{z})\mathbb{E}(\boldsymbol{x}|\boldsymbol{z})^\dagger)\mathbf{B}^\dagger\mathbf{G}^\dagger$$

Therefore,

$$\frac{\partial I(\boldsymbol{x};\boldsymbol{z})}{\partial \mathbf{A}^\dagger} = \mathbf{M}\mathbb{E}(\boldsymbol{x}\boldsymbol{x}^\dagger)\mathbf{B}^\dagger\mathbf{G}^\dagger - \mathbf{M}\mathbb{E}(\mathbb{E}(\boldsymbol{x}|\boldsymbol{z})\mathbb{E}(\boldsymbol{x}|\boldsymbol{z})^\dagger)\mathbf{B}^\dagger\mathbf{G}^\dagger$$
$$\frac{\partial I(\boldsymbol{x};\boldsymbol{z})}{\partial \mathbf{A}^\dagger} = \mathbf{MEB}^\dagger\mathbf{G}^\dagger$$

Therefore, Theorem 1-Part I has been proved.

## APPENDIX B

### Proof of Theorem 1-Part II

The derivative of the conditional output can be written as,

$$\frac{\partial p_{z|x}(\boldsymbol{z}|\boldsymbol{x})}{\partial \mathbf{G}^\dagger} = -p_{z|x}(\boldsymbol{z}|\boldsymbol{x})\frac{\partial}{\partial \mathbf{G}^\dagger}(\boldsymbol{z}-\mathbf{M}\boldsymbol{x})^\dagger(\boldsymbol{z}-\mathbf{M}\boldsymbol{x})$$
$$= -p_{z|x}(\boldsymbol{z}|\boldsymbol{x})\frac{\partial}{\partial \mathbf{G}^\dagger}(\boldsymbol{z}-\mathbf{AGB}\boldsymbol{x})^\dagger(\boldsymbol{z}-\mathbf{AGB}\boldsymbol{x})$$
$$= p_{z|x}(\boldsymbol{z}|\boldsymbol{x})(\boldsymbol{z}-\mathbf{M}\boldsymbol{x})\mathbf{A}^\dagger\boldsymbol{x}^\dagger\mathbf{B}^\dagger$$
$$= \nabla_z p_{z|x}(\boldsymbol{z}|\boldsymbol{x})\mathbf{A}^\dagger\boldsymbol{x}^\dagger\mathbf{B}^\dagger$$

Therefore we have,

$$\mathbb{E}_x\left(\nabla_{A^\dagger} p_{z|x}(\boldsymbol{z}|\boldsymbol{x})\right) = \mathbb{E}_x\left(\nabla_z p_{z|x}(\boldsymbol{z}|\boldsymbol{x})\mathbf{A}^\dagger\boldsymbol{x}^\dagger\mathbf{B}^\dagger\right)$$
$$\frac{\partial I(\boldsymbol{x};\boldsymbol{z})}{\partial \mathbf{G}^\dagger} = \int (1 + log\, p(\boldsymbol{z}))\,\mathbb{E}_x\left(\nabla_z p_{z|x}(\boldsymbol{z}|\boldsymbol{x})\right)\mathbf{A}^\dagger\boldsymbol{x}^\dagger\mathbf{B}^\dagger d\boldsymbol{z}$$
$$= \mathbb{E}\left(\left(\int (1 + log\, p(\boldsymbol{z}))\,\nabla_z p_{z|x}(\boldsymbol{z}|\boldsymbol{x}) d\boldsymbol{z}\right)\mathbf{A}^\dagger\boldsymbol{x}^\dagger\mathbf{B}^\dagger\right)$$

Using integration by parts applied to the real and imaginary parts of $\boldsymbol{z}$,

$$\int (1 + log\ _e p(\mathbf{z})) \frac{\partial p_{z|x}(\mathbf{z}|\boldsymbol{x})}{\partial \boldsymbol{t}} dt$$

$$= \int (1 + log\ _e p(\mathbf{z}))\, p_{z|x}(\mathbf{z}|\boldsymbol{x})|_{-\infty}^{\infty} - \int_{-\infty}^{\infty} \frac{1}{p(\mathbf{z})} \frac{\partial p(\mathbf{z})}{\partial \boldsymbol{t}}\, p_{z|x}(\mathbf{z}|\boldsymbol{x}) dt$$

The first term goes to zero as $\|\mathbf{z}\| \xrightarrow{yields} \infty$, Then,

$$\frac{\partial I(\boldsymbol{x};\mathbf{z})}{\partial \mathbf{G}^\dagger} = \mathbb{E}\left(\left(-\int \frac{p_{z|x}(\mathbf{z}|\boldsymbol{x})}{p(\mathbf{z})} \nabla_z p(\mathbf{z})\, d\mathbf{z}\right) \mathbf{A}^\dagger \boldsymbol{x}^\dagger \mathbf{B}^\dagger\right)$$

$$= -\int \nabla_z p(\mathbf{z}) \mathbb{E}_x \left(\frac{p_{z|x}(\mathbf{z}|\boldsymbol{x})}{p(\mathbf{z})} \mathbf{A}^\dagger \boldsymbol{x}^\dagger \mathbf{B}^\dagger\right) d\mathbf{z}$$

$$= -\int \nabla_z p(\mathbf{z})\, \mathbf{A}^\dagger \mathbb{E}(\boldsymbol{x}|\mathbf{z})^{\dagger}\, \mathbf{B}^\dagger d\mathbf{z}$$

However,

$$\nabla_z p(\mathbf{z}) = \nabla_z \mathbb{E}_x \left(p_{z|x}(\mathbf{z}|\boldsymbol{x})\right) = \mathbb{E}_x \left(\nabla_z p_{z|x}(\mathbf{z}|\boldsymbol{x})\right) = -\mathbb{E}_x \left(p_{z|x}(\mathbf{z}|\boldsymbol{x})(\mathbf{z} - \mathbf{M}\boldsymbol{x})\right)$$

$$= -\mathbb{E}_x(p_z(\mathbf{z})(\mathbf{z} - \mathbf{M}\boldsymbol{x})|\mathbf{z})$$

$$= -p_z(\mathbf{z})\big(\mathbf{z} - \mathbf{M}\mathbb{E}(\boldsymbol{x}|\mathbf{z})\big)$$

Thus,

$$\frac{\partial I(\boldsymbol{x};\mathbf{z})}{\partial \mathbf{G}^\dagger} = \int p_z(\mathbf{z}) \mathbf{A}^\dagger \big(\mathbf{z} - \mathbf{M}\mathbb{E}(\boldsymbol{x}|\mathbf{z})\big)\, \mathbb{E}(\boldsymbol{x}|\mathbf{z})^{\dagger}\, \mathbf{B}^\dagger d\mathbf{z}$$

$$= \mathbf{A}^\dagger \mathbb{E}(z\boldsymbol{x}^\dagger)\mathbf{B}^\dagger - \mathbf{A}^\dagger \mathbb{E}(\mathbf{M}\mathbb{E}(\boldsymbol{x}|\mathbf{z})\mathbb{E}(\boldsymbol{x}|\mathbf{z})^{\dagger})\mathbf{B}^\dagger$$

Therefore,

$$\frac{\partial I(\boldsymbol{x};\mathbf{z})}{\partial \mathbf{G}^\dagger} = \mathbf{A}^\dagger \mathbf{M}\mathbb{E}(\boldsymbol{x}\boldsymbol{x}^\dagger)\mathbf{B}^\dagger - \mathbf{A}^\dagger \mathbf{M}\mathbb{E}(\mathbb{E}(\boldsymbol{x}|\mathbf{z})\mathbb{E}(\boldsymbol{x}|\mathbf{z})^{\dagger})\mathbf{B}^\dagger$$

$$\frac{\partial I(\boldsymbol{x};\mathbf{z})}{\partial \mathbf{G}^\dagger} = \mathbf{A}^\dagger\ \mathbf{M}\mathbf{E}\mathbf{B}^\dagger$$

Therefore, Theorem 1-Part II has been proved.

## APPENDIX C

### Proof of Theorem 1- Part III

The derivative of the conditional output can be written as,

$$\frac{\partial p_{z|x}(\mathbf{z}|\boldsymbol{x})}{\partial \mathbf{B}^\dagger} = -p_{z|x}(\mathbf{z}|\boldsymbol{x}) \frac{\partial}{\partial \mathbf{B}^\dagger} (\mathbf{z} - \mathbf{M}\boldsymbol{x})^\dagger (\mathbf{z} - \mathbf{M}\boldsymbol{x})$$

$$= -p_{z|x}(\mathbf{z}|\boldsymbol{x}) \frac{\partial}{\partial \mathbf{B}^\dagger} (\mathbf{z} - \mathbf{AGB}\boldsymbol{x})^\dagger (\mathbf{z} - \mathbf{AGB}\boldsymbol{x})$$

$$= p_{z|x}(\mathbf{z}|\boldsymbol{x})(\mathbf{z} - \mathbf{M}\boldsymbol{x})\mathbf{G}^\dagger \mathbf{A}^\dagger \boldsymbol{x}^\dagger$$

$$= \nabla_z p_{z|x}(\mathbf{z}|\boldsymbol{x})\mathbf{G}^\dagger \mathbf{A}^\dagger \boldsymbol{x}^\dagger$$

Therefore we have,

$$\mathbb{E}_x \left(\nabla_{\boldsymbol{A}^\dagger} p_{z|x}(\mathbf{z}|\boldsymbol{x})\right) = \mathbb{E}_x\big(\nabla_z p_{z|x}(\mathbf{z}|\boldsymbol{x})\mathbf{G}^\dagger \mathbf{A}^\dagger \boldsymbol{x}^\dagger\big)$$

$$\frac{\partial I(\boldsymbol{x};\mathbf{z})}{\partial \mathbf{B}^\dagger} = \int (1 + log\ p(\mathbf{z}))\, \mathbb{E}_x \left(\nabla_z p_{z|x}(\mathbf{z}|\boldsymbol{x})\right) \mathbf{G}^\dagger \mathbf{A}^\dagger \boldsymbol{x}^\dagger d\mathbf{z}$$

$$= \mathbb{E}\left(\left(\int (1 + log\ p(\mathbf{z}))\, \nabla_z p_{z|x}(\mathbf{z}|\boldsymbol{x}) d\mathbf{z}\right) \mathbf{G}^\dagger \mathbf{A}^\dagger \boldsymbol{x}^\dagger\right)$$

Using integration by parts applied to the real and imaginary parts of $\mathbf{z}$,

$$\int (1 + log\ _e p(\mathbf{z})) \frac{\partial p_{z|x}(\mathbf{z}|\boldsymbol{x})}{\partial \boldsymbol{t}} dt$$

$$= \int (1 + log\ _e p(\mathbf{z}))\, p_{z|x}(\mathbf{z}|\boldsymbol{x})|_{-\infty}^{\infty} - \int_{-\infty}^{\infty} \frac{1}{p(\mathbf{z})} \frac{\partial p(\mathbf{z})}{\partial \boldsymbol{t}}\, p_{z|x}(\mathbf{z}|\boldsymbol{x}) dt$$

The first term goes to zero as $\|\mathbf{z}\| \xrightarrow{yields} \infty$, Then,

$$\frac{\partial I(\boldsymbol{x};\mathbf{z})}{\partial \mathbf{A}^\dagger} = \mathbb{E}\left(\left(-\int \frac{p_{z|x}(\mathbf{z}|\boldsymbol{x})}{p(\mathbf{z})} \nabla_z p(\mathbf{z})\, d\mathbf{z}\right) \mathbf{G}^\dagger \mathbf{A}^\dagger \boldsymbol{x}^\dagger\right)$$

$$= -\int \nabla_z p(\mathbf{z}) \mathbb{E}_x \left(\frac{p_{z|x}(\mathbf{z}|\boldsymbol{x})}{p(\mathbf{z})} \mathbf{G}^\dagger \mathbf{A}^\dagger \boldsymbol{x}^\dagger\right) d\mathbf{z}$$

$$= -\int \nabla_z p(\mathbf{z})\, \mathbf{G}^\dagger \mathbf{A}^\dagger \mathbb{E}(\boldsymbol{x}|\mathbf{z})^{\dagger}\, d\mathbf{z}$$

However,

$$\nabla_z p(\mathbf{z}) = \nabla_z \mathbb{E}_x \left(p_{z|x}(\mathbf{z}|\boldsymbol{x})\right) = \mathbb{E}_x \left(\nabla_z p_{z|x}(\mathbf{z}|\boldsymbol{x})\right) = -\mathbb{E}_x \left(p_{z|x}(\mathbf{z}|\boldsymbol{x})(\mathbf{z} - \mathbf{M}\boldsymbol{x})\right)$$

$$= -\mathbb{E}_x(p_z(\mathbf{z})(\mathbf{z} - \mathbf{M}\boldsymbol{x})|\mathbf{z})$$

$$= -p_z(\mathbf{z})\big(\mathbf{z} - \mathbf{M}\mathbb{E}(\boldsymbol{x}|\mathbf{z})\big)$$

Thus,

$$\frac{\partial I(\boldsymbol{x};\mathbf{z})}{\partial \mathbf{B}^\dagger} = \int p_z(\mathbf{z}) \mathbf{G}^\dagger \mathbf{A}^\dagger \big(\mathbf{z} - \mathbf{M}\mathbb{E}(\boldsymbol{x}|\mathbf{z})\big)\, \mathbb{E}(\boldsymbol{x}|\mathbf{z})^{\dagger}\, d\mathbf{z}$$

$$= \mathbf{G}^\dagger \mathbf{A}^\dagger \mathbb{E}(z\boldsymbol{x}^\dagger) - \mathbf{G}^\dagger \mathbf{A}^\dagger \mathbb{E}(\mathbf{M}\mathbb{E}(\boldsymbol{x}|\mathbf{z})\mathbb{E}(\boldsymbol{x}|\mathbf{z})^{\dagger})$$

Therefore,

$$\frac{\partial I(\boldsymbol{x};\mathbf{z})}{\partial \mathbf{B}^\dagger} = \mathbf{G}^\dagger \mathbf{A}^\dagger \mathbf{M}\mathbb{E}(\boldsymbol{x}\boldsymbol{x}^\dagger) - \mathbf{G}^\dagger \mathbf{A}^\dagger \mathbf{M}\mathbb{E}(\mathbb{E}(\boldsymbol{x}|\mathbf{z})\mathbb{E}(\boldsymbol{x}|\mathbf{z})^{\dagger})$$

$$\frac{\partial I(\boldsymbol{x};\mathbf{z})}{\partial \mathbf{B}^\dagger} = \mathbf{G}^\dagger \mathbf{A}^\dagger \mathbf{M}\mathbf{E}$$

Therefore, Theorem 1-Part III has been proved.

## APPENDIX D

### Proof of Theorem 2

From the steps in Theorem 1, we can see that,

$$p_z(\mathbf{z})\big(\mathbf{z} - \mathbf{M}\mathbb{E}(\boldsymbol{x}|\mathbf{z})\big) = \mathbb{E}_x \left(p_{z|x}(\mathbf{z}|\boldsymbol{x})(\mathbf{z} - \mathbf{M}\boldsymbol{x})\right)$$

Therefore,

$$\mathbf{M}\mathbb{E}(\boldsymbol{x}|\mathbf{z}) = \boldsymbol{z} - \frac{\mathbb{E}_x \left(p_{z|x}(\mathbf{z}|\boldsymbol{x})(\mathbf{z} - \mathbf{M}\boldsymbol{x})\right)}{p_z(\mathbf{z})}$$

$$= \boldsymbol{z} + \frac{\mathbb{E}_x \left(\nabla_z p_{z|x}(\mathbf{z}|\boldsymbol{x})\right)}{p_z(\mathbf{z})}$$

$$= \boldsymbol{z} + \frac{\nabla_z \mathbb{E}_x \left(p_{z|x}(\mathbf{z}|\boldsymbol{x})\right)}{p_z(\mathbf{z})}$$

Thus,

$$\mathbf{M}\mathbb{E}(\boldsymbol{x}|\mathbf{z}) = \boldsymbol{z} + \frac{\nabla_z p_z(\mathbf{z})}{p_z(\mathbf{z})}$$

Then we can write the input estimates as,

$$\mathbb{E}(\boldsymbol{x}|\mathbf{z}) = \left[\boldsymbol{x} + \frac{\nabla_z p_z(\mathbf{z})}{p_z(\mathbf{z})}\right] \mathbf{M}^{-1}$$

$$= [\boldsymbol{x} + \nabla_z \log p_z(\mathbf{z})]\mathbf{M}^{-1}$$

$$= [\boldsymbol{x} + \nabla_z \log p_z(\mathbf{z})](\mathbf{AGB})^{-1}$$

$$= [\boldsymbol{x} + \nabla_z \log p_z(\mathbf{z})]\mathbf{B}^{-1}\mathbf{G}^{-1}\mathbf{A}^{-1}$$

$$= [\boldsymbol{x} + \nabla_z \log p_z(\mathbf{z})]\mathbf{B}^{-1}(\mathbf{I} - \mathbf{F})\mathbf{A}^{-1}$$

Therefore, Theorem 2 has been proved.

## REFERENCES

[1] D. Guo, S. Shamai, and S. Verdú, "Mutual information and minimum mean-square error in Gaussian channels," *IEEE Trans. Inf. Theory*, vol. 51, no. 4, pp. 1261–1282, Apr. 2005.

[2] D. P. Palomar and S. Verdú, "Gradient of mutual information in linear vector Gaussian channels," *IEEE Trans. Inf. Theory*, vol. 52, no. 1, pp. 141–154, Jan. 2006.

[3] S. A. M. Ghanem, "MAC Gaussian channels with arbitrary inputs: optimal precoding and power allocation," *IEEE International Conference on Wireless Communications and Signal Processing (WCSP),* Oct. 2012.

[4] S. A. M. Ghanem, "Multiple access Gaussian channels with arbitrary inputs: Optimal Precoding and Power Allocation," available on ArXiv, 2014.

[5] S. A. M. Ghanem, "Mutual information and minimum mean-sqare error in multiuser Gaussian channels," invited to *IEEE MMTC E-Letter, special issue on latest advances on wireless and mobile communications and networking,* Jan, 2015.

[6] S. A. M. Ghanem, "Multiuser I-MMSE'', available on ArXiv, 2015.

[7] T. M. Cover and J. A. Thomas, Elements of Information Theory, 2nd ed. New York: Wiley, 2006.

[8] R. Koetter, M. Médard, "An Algebraic Approach to Network Coding," *IEEE/ACM Trans. On Networking,* vol. 11, no. 5, Oct. 2003.

[9] R. Alswede, "Network Information Flow," *IEEE Trans. on Inf. Theory,* vol. 46, no. 4, July, 2000.